\documentclass[sigconf, nonacm]{acmart}

\usepackage{booktabs} 

\usepackage[utf8]{inputenc}
\usepackage{enumitem}
\usepackage{xcolor}
\usepackage{amsmath}
\usepackage{subfigure}
\usepackage{cleveref}
\usepackage{listings}
\usepackage{fancyvrb}
\usepackage{xurl}
\usepackage{xspace}
\definecolor{darkgray}{rgb}{0.33, 0.33, 0.33}

\usepackage{balance}

\newcommand{\header}[1]{\vspace{1mm}\noindent\textbf{#1}.}
\newcommand{\headerl}[1]{\vspace{1mm}\noindent\textit{#1}.}

\newcommand{\mlinspect}{\texttt{mlinspect}\xspace}
\newcommand{\mlwhatif}{\texttt{mlwhatif}\xspace}

\newcommand\vldbpagestyle{plain} 

\begin{document}
\title{Instrumentation and Analysis of Native \\ML Pipelines via Logical Query Plans}

\author{Stefan Grafberger}
\email{grafberger@tu-berlin.de}
\orcid{0000-0002-9884-9517}
\affiliation{%
    Supervised by Sebastian Schelter and Paul Groth\\
  \institution{BIFOLD \& TU Berlin}
  \country{Germany}
}

\begin{abstract}
Machine Learning (ML) is increasingly used to automate impactful decisions, which leads to concerns regarding their correctness, reliability, and fairness. We envision highly-automated software platforms to assist data scientists with developing, validating, monitoring, and analysing their ML pipelines. In contrast to existing work, our key idea is to extract ``logical query plans'' from ML pipeline code relying on popular libraries. Based on these plans, we automatically infer pipeline semantics and instrument and rewrite the ML pipelines to enable diverse use cases without requiring data scientists to manually annotate or rewrite their code.

First, we developed such an abstract ML pipeline representation together with machinery to extract it from Python code. Next, we used this representation to efficiently instrument static ML pipelines and apply provenance tracking, which enables lightweight screening for common data preparation issues. Finally, we built machinery to automatically rewrite ML pipelines to perform more advanced what-if analyses and proposed using multi-query optimisation for the resulting workloads. In future work, we aim to interactively assist data scientists as they work on their ML pipelines.

\end{abstract}

\maketitle

\pagestyle{\vldbpagestyle}

\section{introduction}
\label{sec:introduction}
\begin{figure*}[t]
  \centering
  \includegraphics[scale=0.3]{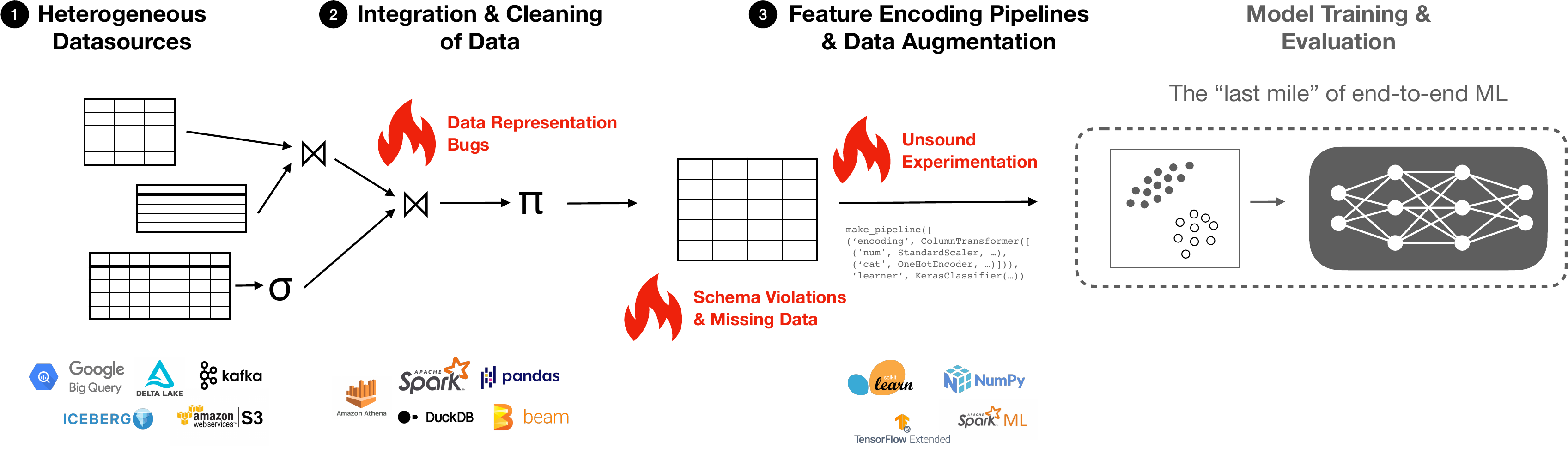}
  \caption{\textbf{ML Pipelines in the real world often join data from multiple data sources, clean and integrate the data, define feature encoding pipelines, and use techniques like data augmentation before finally passing the featurised data to ML models. The model training and evaluation, which is typically the focus of ML research, is only a small part of the process.}}
  \label{fig:ml-pipelines-real-world}
\end{figure*}

Machine Learning (ML) is increasingly used to automate critical decisions in domains like credit and lending, medical diagnosis, and hiring, with the potential to reduce costs, reduce errors, and make outcomes more equitable~\cite{julia2022}. Yet, despite its potential, the risks arising from the widespread use of ML are garnering attention from policymakers, scientists, and the media~\cite{julia2022}. In large part, this is because the correctness, reliability, and fairness of ML models critically depend on their training data.
Data quality issues, pre-existing bias, such as under- or over-representation of particular groups in the training data
, and technical bias, such as skew introduced during data preparation, can heavily impact performance. Furthermore, creating reliable and robust ML pipelines requires a lot of expertise in various areas like ML, MLOps, and software engineering.
As shown in Figure~\ref{fig:ml-pipelines-real-world}, the input data for such ML applications has to be integrated, preprocessed, and cleaned first. The data preparation part before the ML model also heavily impacts the pipeline performance~\cite{mazumder2023dataperf}, motivating our research. 

\header{Gap between ML research and ML usage in industry}
In many cases, there is a \textit{gap between ML research and ML usage in industry}. Usually, existing research assumes a single, clean, fully-integrated, static dataset that is ready to be featurized and fed to an ML model using a simple Juptyer notebook, written by ML and statistics experts. Meanwhile, in the real world, practitioners often spend large portions of their time on tasks like data loading, data cleaning, and model deployment~\cite{anaconda}. Many of them may have a background in software engineering rather than in ML and statistics. 

\header{Missing theoretical foundation for end-to-end ML pipelines}
Detecting many issues in ML pipelines, such as data distribution bugs and skew during data preparation, is challenging because \textit{different pipeline steps are implemented using different libraries and abstractions, and the data representation often changes from relational data to matrices during data preparation}. Further, preprocessing in the data science ecosystem~\cite{dataScienceLookingGlass} 
often combines relational operations on tabular data with  \textit{estimator/transformer pipelines}~\cite{estimatorTransformer}, a composable and nestable abstraction for operations on array data. 
Without an abstract pipeline representation that works across the boundaries of libraries, reasoning about pipelines is difficult.

\header{Automatically understanding ML pipeline semantics}
Due to the pressures of their day-to-day activities, \textit{most data scientists cannot invest the time and effort to manually instrument their code or insert logging statements for tracing}. 
Furthermore, currently established data science libraries are here to stay in the foreseeable future (due to the high investments made in them already), so replacing them with new holistic frameworks is also unrealistic. Instead, we envision directly using existing pipeline code as input. 

\header{Logical query plans for ML pipelines}
To achieve this, we propose to extract ``logical query plans'' from ML pipelines. 
We can then leverage these plans to automatically instrument and rewrite ML code to enable different use cases. In particular, we focus on {\em natively written} ML pipelines that use established libraries from the data science ecosystem, such as pandas, scikit-learn or keras. 
We do not require data scientists to manually annotate or rewrite their code. However, this approach relies on declaratively written ML pipelines, where we can identify the semantics of the operations.

\header{Achieved and planned contributions}
Abstractions like logical query plans are the foundation for modern databases. We believe that corresponding abstractions are also fundamental for better software to assist data scientists with creating robust, reliable, and fair ML pipelines. We envision highly automated platforms to assist with tasks like developing, validating, monitoring, and analysing ML pipelines.
During my Ph.D., we started working towards this vision. First, we developed an ML pipeline representation and methods for extracting the resulting plans from Python code. Following that, we built a light-weight library called \mlinspect{}~\cite{grafberger2021lightweight,grafberger2021demo,grafberger2022data}, which uses these plans to efficiently instrument static ML pipelines, transparently observe their execution, and apply record-level provenance tracking. Building on this runtime, we show how to apply it as a foundation for tasks like lightweight screening of common data preparation issues.  In our next project, \mlwhatif{}~\cite{grafberger2022towards,grafberger2023mlwhatif,grafberger2023demo}, we built machinery to automatically rewrite pipelines to perform more advanced what-if analyses and proposed using multi-query optimisation for the resulting workloads. For the final PhD project, we aim to interactively assist data scientists working on ML pipelines~\cite{grafberger2024towards}.

In summary, we achieved the following contributions so far:
\begin{itemize}[noitemsep,leftmargin=*]
    \item We define an abstract representation for ML pipelines and propose machinery to extract the resulting plans from ML pipelines in Python using popular libraries~(\Cref{sec:dag}).
    \item We propose methods to efficiently instrument static ML pipelines and use provenance-tracking for light-weight screening for common data preparation issues~(\Cref{sec:mlinspect}). 
    \item We propose methods for dynamically rewriting ML pipelines to perform more advanced what-if analyses and optimising the resulting workloads via multi-query optimisation~(\Cref{sec:mlwhatif}). 
    \item We propose initial ideas to interactively assist data scientists with improving ML data preparation code~(\Cref{sec:mlidea}).
\end{itemize}
\section{Achieved Results}
After defining the abstract pipeline representation, we used it as a foundation for two projects: light-weight inspection of ML pipelines~\cite{grafberger2021lightweight,grafberger2021demo,grafberger2022data} and data-centric what-if analysis~\cite{grafberger2022towards,grafberger2023mlwhatif,grafberger2023demo}.
\subsection{Logical Query Plans for ML Pipelines}
\label{sec:dag}

\begin{figure*}[t]
  \centering
  \includegraphics[scale=0.57]{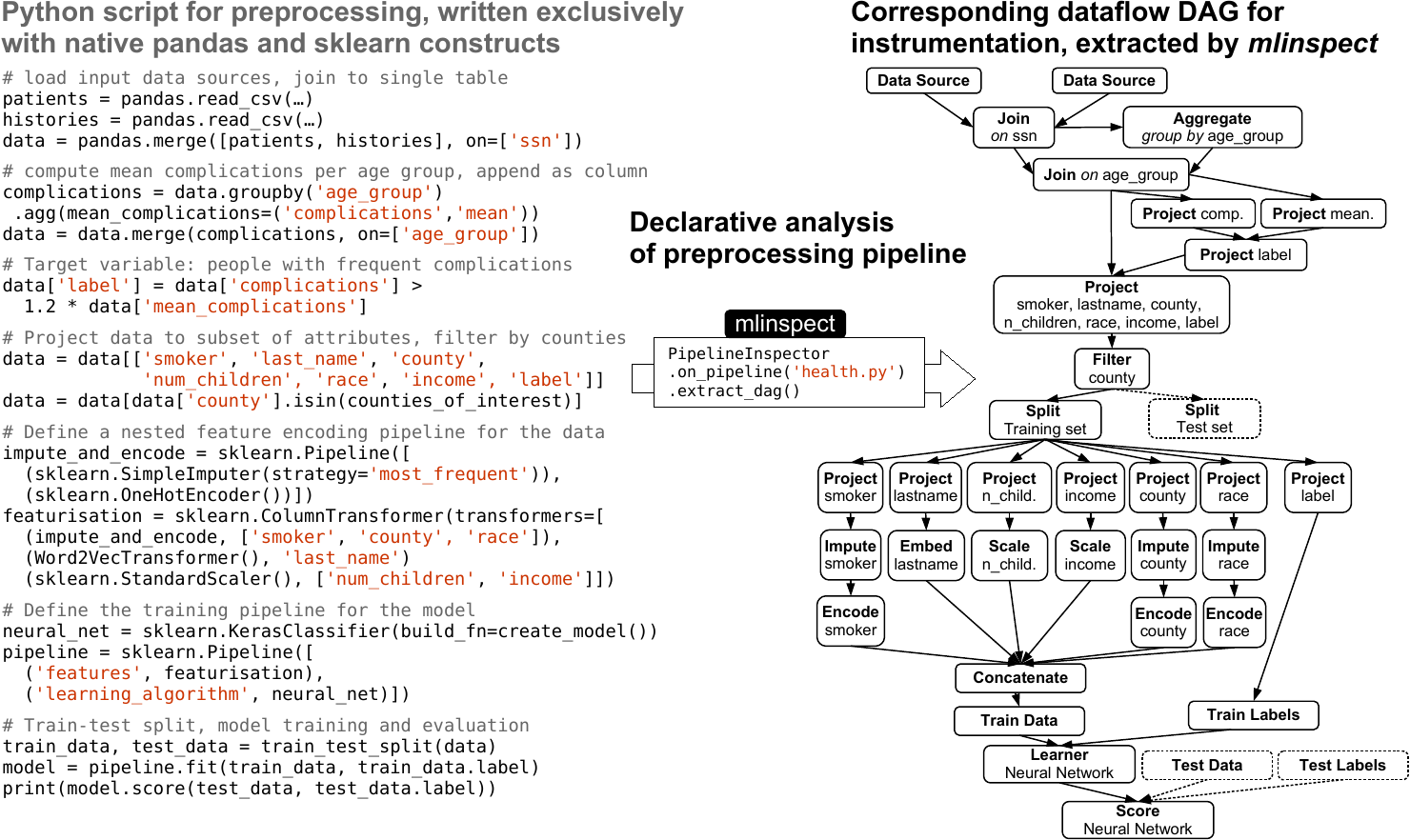}
  \caption{\textbf{Example of an ML pipeline in healthcare that predicts which patients are at a higher risk of serious medical complications. The pipeline is implemented using native constructs from the popular pandas and scikit-learn libraries. On the left, we show the source code of the pipeline. On the right, we show the corresponding dataflow graph extracted by our methods. (Operations on the test set and for estimator/transformer fitting are omitted for readability.)}}
  \label{fig:dag-example}
\end{figure*}

In database systems, logical query plans are not only the central abstraction for query optimisation, but also the theoretical foundation for concepts like provenance~\cite{green17semiringFramework}, what-if analysis~\cite{deutch2013caravan}, incremental view maintenance~(IVM)~\cite{mcsherry2013differential}, and techniques for explaining query outputs~\cite{sudeepa15explainingAnswers}. Similarly, we propose using logical query plans for ML pipelines as a foundation to develop comparable techniques for tasks like developing, validating, monitoring, and analysing ML pipelines. Figure~\ref{fig:dag-example} shows such a plan for an exemplary ML pipeline.

\header{Structure of ML pipelines} In general, ML applications for supervised learning in real-world scenarios work with several input data sources $\mathcal{D}_1, \dots, \mathcal{D}_n$, such as log files, database tables, or files \newpage \noindent in a data lake, often in the form of relational data accompanied by unstructured data such as text and images
. Therefore, ML pipelines consist of three subsequent high-level stages:

\begin{enumerate}[noitemsep,leftmargin=*]
  \item \headerl{Relational preprocessing} This stage integrates the input datasets $\mathcal{D}_1, \dots, \mathcal{D}_n$ into a single training relation $\mathcal{D}_{\text{train}}$ and a single test relation $\mathcal{D}_{\text{test}}$. This stage involves steps like relational joins, data integration, cleaning, filtering, and attribute computation.
  \item \headerl{Featurization} This stages encodes the relations $\mathcal{D}_{\text{train}}$ and $\mathcal{D}_{\text{test}}$ into matrix form, producing the train set $(\mathbf{X}_{\text{train}}, \mathbf{y}_{\text{train}})$ and test set $(\mathbf{X}_{\text{test}}, \mathbf{y}_{\text{test}})$ for the ML model. This stage typically applies ML-specific feature encoding steps based on linear algebra, e.g., one-hot-encoding, embedding, and feature hashing. 
  \item \headerl{Model training and scoring} The final stage conducts the model training based on the train set $(\mathbf{X}_{\text{train}}, \mathbf{y}_{\text{train}})$, producing the model $f_\theta$, and computes the output score $U(f_\theta(\mathbf{X}_{\text{test}}), \mathbf{y}_{\text{test}})$ denoting its prediction quality on the unseen test set.
\end{enumerate}

\header{Estimator/transformers} Furthermore, ML pipelines contain so-called estimator/transformer operations, which are popular in common ML libraries such as scikit-learn, SparkML, Tensorflow Transform, and Ray. These operations are a composable and nestable way to hide the complexity of featurization operations. The estimator part is typically applied to training data, where it conducts a global aggregation to compute statistics, which are then used by a subsequent transformer (a tuple-at-a-time operation) to transform tuples in both train and test data. 

\header{Modeling ML pipelines as dataflow computations} 
The dataflow computation consumes the input datasets $\mathcal{D}_1, \dots, \mathcal{D}_n$ and produces the score $U(f_\theta(\mathbf{X}_{\text{test}}), \mathbf{y}_{\text{test}})$ as output. 
The operators in the dataflow computation correspond to relational operations, e.g., to selections, projections and joins in the relational preprocessing stage. 
We treat ML-specific operations in the featurization stage as global aggregations and extended projections which output arrays, model training as a black-box aggregation, and model prediction again as extended \newpage \noindent projection. Formally, the pipeline is thereby treated as a directed acyclic graph (DAG) whose vertices $V$ correspond to operators for the discussed relational operations, and whose edges correspond to data exchange between the operators.

\subsection{Lighweight Issue Detection via Provenance}
\label{sec:mlinspect}
We then used these logical query plans to develop \mlinspect{}~\cite{grafberger2021lightweight,grafberger2021demo,grafberger2022data}, a library that helps diagnose and mitigate technical bias that may arise during preprocessing steps in an ML pipeline. We refer to these problems collectively as \emph{data distribution bugs}.

\header{Lightweight pipeline inspection with \mlinspect{}} 
After extracting these plans from Python code using popular libraries, \mlinspect{} automatically instruments the code to trace the impact of operators on properties like the distribution of sensitive groups in the data. Via a simple declarative interface, data scientists can automatically check their pipelines for data distribution bugs. 

Importantly, \mlinspect{} provides a library-independent interface to propagate annotations such as the lineage of tuples across operators from different libraries, and introduces only constant overhead per tuple flowing through the DAG. Thereby, \mlinspect{} offers a general runtime for light-weight pipeline inspection, and enables the integration of many detection techniques for data distribution bugs that previously required custom code, such as automated model validation of data slices, identification of distortions with respect to protected group membership in the training data, and automated dataset sanity checking.

\subsection{Data-centric What-if Analysis}
\label{sec:mlwhatif}
Next, we focused on the problem of data-centric what-if analysis for native ML pipelines. We developed \mlwhatif{}~\cite{grafberger2022towards,grafberger2023mlwhatif,grafberger2023demo}, a library to automatically rewrite ML pipelines to test what-if scenarios and optimise the execution of the resulting what-if scenario workloads. 

\header{Data centric what-if analyses on ML pipelines with \mlwhatif{}} During the development of ML pipelines, an important task of data scientists is to understand the sensitivity of their pipeline by performing {\em data-centric what-if analyses}~\cite{grafberger2022towards}. Such analyses, for example, focus on $(i)$ the robustness against data errors, asking {\em what-if the input data to a pipeline had certain errors like missing values or outliers?} $(ii)$ feature importance, asking {\em what-if the pipeline did not have access to a particular feature?}, and $(iii)$ the impact of preprocessing operators on the pipeline's fairness, asking {\em what-if the pipeline cleaned, filtered or featurized the training data differently?} 
These what-if analyses follow a common pattern: they take an existing ML pipeline, create a \textit{pipeline variant} by introducing a small change, and \textit{execute this pipeline variant} to see how the change impacts the pipeline's output score, e.g., its accuracy or a fairness metric.
The application of these analysis techniques to ML pipelines poses several technical challenges.

\header{Integration challenge} Data-centric what-if analyses are difficult to integrate with existing pipeline code. Many existing techniques are designed for single input datasets in matrix form and not for pipelines with multiple heterogeneous input datasets. Additionally, they are often implemented as stand-alone software packages with hardcoded data preparation steps. 
As a result, the integration of such analyses requires significant and costly manual development efforts. Reducing this development time is crucial, as data scientists already spent more than 60\% of their time on data preparation tasks~\cite{anaconda}. 
To address this, we developed ``pipeline patches'' as a formal framework to generate different variants of an ML pipeline, specifying changes to their input data, operators, and models. 

\header{Efficiency challenge} A major part of the computation time in ML pipelines is spent on data  preparation and validation. This overhead grows when we run what-if analyses on a pipeline, as the repeated execution of the pipeline variants incurs a lot of redundant work. Thus, we propose using multi-query optimization for the joint execution of several pipeline variants to re-use shared intermediates to reduce the runtime of what-if analyses.

\section{Future Directions}
\label{sec:mlidea}
In the remainder of the PhD, we intend to work on interactively improving ML pipeline code~\cite{grafberger2024towards}.

\header{Previous work}
Over time, more and more data-centric techniques are being developed to detect, quantify, and improve ML applications with respect to their reliability, fairness, and prediction quality~\cite{mazumder2023dataperf,grafberger2023mlwhatif}.  
However, applying these techniques to ML pipelines still requires a high level of expertise, as previous approaches assume that data scientists know in advance what errors they are looking for. 

\header{The need for interactively improving ML pipelines} In reality, data scientists typically do not know in advance what pipeline issues to look for, and often ``discover serious issues only after deploying their systems in the real world''~\cite{holstein2019improving}. At development time, data scientists currently have to iteratively screen their pipeline for potential issues, debug these issues, and then revise and improve the pipelines according to their findings. This process is tedious, as it requires repeated manual code re-organisation and re-execution in an environment like a Jupyter notebook.

We argue that ML pipeline development should be accompanied by \textit{interactive suggestions} to improve the pipeline code, similar to code inspections in modern IDEs like IntelliJ~\cite{codeInspectionsIntelliJ} or text corrections in writing assistants like Grammarly~\cite{grammarlyDemo}. For that, we can re-use some techniques from previous work, but are still faced with a set of challenges:
$(i)$~We need \textit{low-latency auto-detection} of pipeline improvement opportunities, to seamlessly integrate into the development workflow; $(ii)$ we should identify \textit{pipeline problems spanning several operators}, instead of being artificially limited to screening individual operators one-at-a-time as in e.g., \mlinspect{}~\cite{grafberger2021lightweight,grafberger2021demo,grafberger2022data}; also, $(iii)$ users should receive~\textit{provenance-enabled explanations} for detected problems and suggested improvements.

\header{Low-latency suggestions for improvement} We envision a system that instruments a data scientist's ML pipeline code and creates and maintains so-called ``\textit{shadow pipelines}'' with low-latency to generate suggestions for improvements. 
Such a shadow pipeline is a hidden variant of the original pipeline, which modifies it to auto-detect potential issues and tries out different pipeline modifications for improvement opportunities. Subsequently, each shadow pipeline provides the user with code suggestions to improve the pipeline, accompanied by a provenance-based explanation and a \newpage \noindent quantification of the expected impact on the pipeline outputs. From a technical perspective, the main challenge is to conduct the required computations with low latency by reusing and updating intermediates via incremental view maintenance. 

\section{Conclusion}
\label{sec:conclusion}
We outlined my PhD research agenda, which uses an abstract representation of ML pipelines as the foundation to partially automate their development, validation, and analysis. Just as modern databases are built on abstractions like logical query plans, we aim to show that a corresponding theoretical foundation is fundamental for better data science tooling. During my PhD, we already showed how these plans can be used as foundation for light-weight screening for common data preparation issues~\cite{grafberger2021lightweight,grafberger2021demo,grafberger2022data} and automatically performing data-centric what-if analyses~\cite{grafberger2022towards,grafberger2023mlwhatif,grafberger2023demo}. Next, we are planning to interactively support data scientists with improving ML pipelines~\cite{grafberger2024towards}. All of this research is or will be open-source~\cite{linkMlinspect,linkMlwhatif}.

We believe these ideas can benefit many more use cases in the future, e.g., experiment tracking~\cite{schelter2022screening,schelter2023demo,grafberger24redOnions}, CI tooling to proactively screen ML pipelines for issues~\cite{schelter2022screening,schelter2023demo}, auditing~\cite{grafberger24redOnions}, production monitoring~\cite{grafberger24redOnions}, and the simplification of ML pipeline debugging.

\bibliographystyle{ACM-Reference-Format}
\bibliography{main}

\end{document}